\documentclass[journal,10pt]{IEEEtran}

\usepackage{color}
\usepackage{multirow}
\usepackage{graphicx}
\usepackage{epstopdf}
\usepackage[cmex10]{amsmath}
\usepackage{amssymb}

\usepackage{multirow}
\usepackage{amsthm}
\usepackage{cite}
\usepackage{acronym} 
\usepackage{algorithm, algcompatible}

\newcommand{\diag}{\mathop{\rm diag}}

\algnewcommand\INPUT{\item[\textbf{Input:}]}
\algnewcommand\OUTPUT{\item[\textbf{Output:}]}

\allowdisplaybreaks

\begin{document}

\title{Zero Forcing Detection For Short Packet Transmission Under Channel Estimation Errors} 
\author{Nikolaos~I.~Miridakis, Theodoros~A.~Tsiftsis,~\IEEEmembership{Senior Member,~IEEE}, and Hui-Ming~Wang,~\IEEEmembership{Senior Member,~IEEE}
\thanks{N. I. Miridakis and T. A. Tsiftsis are with the School of Electrical and Information Engineering and the Institute of Physical Internet, Jinan University, Zhuhai 519070, China. N. I. Miridakis is also with the Department of Electrical and Electronic Engineering, University of West Attica, Aegaleo 12244, Greece (e-mails: nikozm@uniwa.gr, theo\_tsiftsis@jnu.edu.cn).}
\thanks{H.-M. Wang is with the School of Electronic and Information Engineering, Xi'an Jiaotong University and also with the Ministry of Education Key Laboratory for Intelligent Networks and Network Security, Xi'an, Shaanxi 710049, P. R. China (e-mail: xjbswhm@gmail.com).}
}


\maketitle

\begin{abstract}
A multiuser multiple-input multiple-output wireless communication system is analytically studied under the short-packet transmission regime. The practical scenario of channel estimation errors is adopted when the signals undergo Rayleigh channel fading conditions. Unlike most previous works, the channel estimation error term is treated as a signal rather than interference or noise, which may further enhance the overall system performance. The spatial multiplexing mode of transmit operation is considered, while the zero-forcing detection is applied at the receiver. New and exact closed-form expressions for the outage probability and total system goodput are derived. Capitalizing on the analytical results, some new engineering insights are also presented; such as the impact of channel estimation errors with respect to the number of antennas, transmit power, and/or coding rate.
\end{abstract}


\IEEEpeerreviewmaketitle

\section{Introduction}
\IEEEPARstart{M}{achine} type and ultra-reliable low-latency communications (URLLC) are already indispensable use cases of fifth Generation (5G) networks. Ultra-high reliability of even more than $99.9999\%$ and low latency in the order of less than $1$ms represent two of the cornerstone prerequisites of the latter cases with applications ranging from remote surgery to virtual reality. Unlike conventional communication realized with large packets approaching Shannon's infinite blocklength assumption, the 5G low latency applications can only be realized in the finite blocklength regime (few hundreds of channel uses; e.g., $100$-$300$ for vehicle-to-vehicle communication \cite{j:MakkiSvensson2014}). The performance of short-packet communications has lately attracted significant research interest. However, most of the works assume only perfect channel-state information (CSI) conditions (e.g., see \cite{j:MakkiSvensson2014,j:MakkiSvenssonColdreyy2019,j:MakkiSvenssonn2019,j:HuSchmeinkk2016,j:HuGrosss2015,j:HuGrossss2016,c:GuVucetic2018} and relevant references therein). Nevertheless, perfect CSI is a rather overoptimistic condition in realistic scenarios, mainly due to the unexpected user mobility, vast channel fading variations, and lack of feedback signaling. Only recently, imperfect CSI conditions were considered in \cite{j:HuSchmeink2018,j:Schiessl2016}; yet, these studies focused only on single-antenna transceivers.

Motivated by the lack of analytical investigation on the impact of practical CSI estimation schemes on the performance of short-packet multiple-input multiple-output (MIMO) communication, we study spatial multiplexed multiuser MIMO systems operating over Rayleigh faded channels and zero-forcing (ZF) detection at the receiver side. A training phase occurs prior to the data communication phase for CSI acquisition. Unlike most previous works, the channel estimation error term is treated as a signal rather than interference or noise, which may improve the received signal-to-noise ratio (SNR) and, hence, the system performance \cite{j:LiLin2016}. Novel exact closed-form expressions are derived for some key performance metrics; namely, the outage probability and total system goodput. Along with numerically evaluated performance results, some useful engineering insights are provided.

{\it Notation:} Matrices are represented by uppercase bold typeface letters, whereas $\mathbf{I}_{v}$ stands for the $v\times v$ identity matrix. A diagonal matrix with entries $x_{1},\cdots,x_{n}$ is defined as $\diag\{x_{i}\}^{n}_{i=1}$, while $\mathbf{X}^{-1}$ is the inverse of $\mathbf{X}$. Superscript $(\cdot)^{\mathcal{H}}$ denotes Hermitian transposition and $|\cdot|$ represents absolute (scalar) value. $\mathbb{E}[\cdot]$ is the expectation operator and symbol $\overset{\text{d}}=$ means equality in distribution. $f_{X}(\cdot)$ and $F_{X}(\cdot)$ represent the probability density function (PDF) and cumulative distribution function (CDF) of a random variable (RV) $X$, respectively. $\mathcal{CN}(\mu,\sigma^{2})$ and $\mathcal{N}(\mu,\sigma^{2})$ define, respectively, a complex-valued and real-valued Gaussian RV with mean $\mu$ and variance $\sigma^{2}$, and $\mathcal{X}^{2}_{v}(u,w)$ denotes that $X$ is a non-central chi-squared RV with $v$ degrees-of-freedom (DoF), a non-centrality parameter $u$ and variance $w$. $Q(\cdot)$ is the Gaussian $Q$-function; $Q_{1}(\cdot,\cdot)$ is the first-order Marcum-$Q$ function; $I_{n}(\cdot)$ represents the $n$th order modified Bessel function of the first kind \cite[Eq. (8.445)]{tables}; and ${}_1F_{1}(\cdot,\cdot;\cdot)$ is the Kummer's confluent hypergeometric function \cite[Eq. (9.210.1)]{tables}. Also, $(\cdot)_{p}$ is the Pochhammer symbol with $p \in \mathbb{N}$ \cite[p. xliii]{tables}, $\Gamma(\cdot,\cdot)$ denotes the upper incomplete Gamma function \cite[Eq. (8.350.2)]{tables}, and $\Gamma(\cdot)$ denotes the Gamma function \cite[Eq. (8.310.1)]{tables}. Finally, $\mathcal{O}(\cdot)$ is the Landau symbol.

\section{System and Signal Model}
Consider a wireless communication system with $M$ single-antenna transmitters and a receiver equipped with $N\geq M$ antennas\footnote{From the analysis presented hereafter, the classical single-user communication scenario with $M$ co-located transmit antennas is included as a special case.} operating over a quasi-static block-fading channel. In fact, Rayleigh fading conditions are assumed, which remain fixed for the duration of a given frame transmission, while they may change independently amongst various frames. A training phase occurs prior to the actual data transmission phase at each consecutive frame so as to perform channel estimation at the receiver side. This is implemented through pilot symbol assisted modulation (PSAM), while full-blind transmitters are assumed. Each frame/block has a total capacity of $L_{\rm T}\triangleq m_{\rm T}+L$ channel uses, where $m_{\rm T}$ and $L$ denote the channel uses for the training and data communication phase, respectively. After the training phase, the system enters the data transmission phase. In the current work, the spatial multiplexing mode of operation is applied, where $M$ independent data streams are simultaneously transmitted by the corresponding nodes. The suboptimal yet efficient linear ZF detection scheme is applied at the receiver regarding the data communication phase.

More specifically, at the end of the training phase, the received signal reads as $\mathbf{Y}=\sqrt{p}\mathbf{H}\mathbf{\Psi}+\mathbf{N}$, where $\mathbf{Y} \in \mathbb{C}^{N \times m_{\rm T}}$, $\mathbf{H}\in \mathbb{C}^{N \times M}$, $\mathbf{\Psi}\in \mathbb{C}^{M \times m_{\rm T}}$, $\mathbf{N}\in \mathbb{C}^{N \times m_{\rm T}}$ denote, respectively, the received signal, channel fading matrix, transmitted pilot symbols, and additive white Gaussian noise (AWGN) matrix. Also, $p$ is the transmitted power. Further, the pilot symbols are mutually orthogonal (i.e., $m_{\rm T}\geq M$ is required), satisfying the unitary property (i.e., $\mathbf{\Psi}^{\mathcal{H}}\mathbf{\Psi}=\mathbf{I}_{M}$), they are fixed and a priori known to the receiver. Moreover, $\mathbf{H}\overset{\text{d}}=\mathcal{CN}(\mathbf{0},\mathbf{\Sigma}^{2}_{h})$, where $\mathbf{\Sigma}^{2}_{h}\triangleq \diag\{\sigma^{2}_{h_{1}},\ldots,\sigma^{2}_{h_{M}}\}$ with $\sigma^{2}_{h_{i}}$ denoting the (known) large-scale channel fading of the $i^{\rm th}$ transmitter, and $\mathbf{N}\overset{\text{d}}=\mathcal{CN}(\mathbf{0},\mathbf{I}_{N})$; i.e., $p$ reflects the input SNR.

Upon the reception of $\mathbf{Y}$, the least-square (LS) channel estimation is adopted\footnote{Generally, the linear minimum mean-squared error estimator outperforms LS, yet it requires expectation over a vast number of channel uses to provide optimality \cite{b:VanTrees68}. Nonetheless, such a case is infeasible in short-packet transmissions.} yielding $\hat{\mathbf{H}}=\mathbf{H}+\mathbf{E}$, where $\mathbf{E}=\mathcal{CN}(\mathbf{0},\sigma^{2}_{e}\mathbf{I}_{N})$ is the channel estimation error matrix with $\sigma^{2}_{e}=M/(m_{\rm T} p)$ and $\hat{\mathbf{H}}=\mathcal{CN}(\mathbf{0},\mathbf{\Sigma}^{2}_{\hat{h}})$ is the channel estimation matrix, while $\mathbf{\Sigma}^{2}_{\hat{h}}=\diag\{\sigma^{2}_{\hat{h}_{1}},\ldots,\sigma^{2}_{\hat{h}_{M}}\}$ and $\sigma^{2}_{\hat{h}_{i}}\triangleq \sigma^{2}_{h_{i}}+\sigma^{2}_{e}$. Focusing on short packet communication and without any loss of generality, hereinafter we assume that $m_{\rm T}=M$, hence $\sigma^{2}_{e}\triangleq 1/p$. It is convenient for the subsequent analysis to introduce the following equivalent channel model as \cite[Eq. (9)]{j:MaJin2007} $\mathbf{H}=\hat{\mathbf{H}}\mathbf{\Sigma}^{2}_{h}(\mathbf{\Sigma}^{2}_{\hat{h}})^{-1}+\mathbf{Z}$, where $\mathbf{Z}=\mathcal{CN}(\mathbf{0},\mathbf{\Sigma}^{2}_{Z})$ with $\mathbf{\Sigma}^{2}_{Z}=\diag\{\sigma^{2}_{Z_{1}},\ldots,\sigma^{2}_{Z_{M}}\}$ and $\sigma^{2}_{Z_{i}}=\sigma^{2}_{h_{i}} \sigma^{2}_{e}/\sigma^{2}_{\hat{h}_{i}}$ (with $1\leq i\leq M$), while $\hat{\mathbf{H}}$ and $\mathbf{Z}$ are mutually independent.

Afterwards, the system enters the data transmission phase and the received signal can be expressed as $\mathbf{y}_{\rm d}=\sqrt{p}\mathbf{H}\mathbf{s}+\mathbf{n}$, where $\mathbf{y}_{\rm d} \in \mathbb{C}^{N \times 1}$, $\mathbf{s}\in \mathbb{C}^{M \times 1}$, $\mathbf{n}\in \mathbb{C}^{N \times 1}$ denote, respectively, the received signal, transmitted signal vector (where $\mathbf{s}^{\mathcal{H}}\mathbf{s}=\mathbf{I}_{M}$), and AWGN at the data phase. ZF detection via the efficient QR decomposition \cite{j:MiridakisKaragiannidis2014} is applied at the receiver and the detected signal becomes
\begin{align}
\nonumber
\mathbf{r}=\hat{\mathbf{Q}}^{\mathcal{H}}\mathbf{y}_{\rm d}&=\sqrt{p}\hat{\mathbf{Q}}^{\mathcal{H}}\left[\left(\hat{\mathbf{H}}\mathbf{\Sigma}^{2}_{h}\left(\mathbf{\Sigma}^{2}_{\hat{h}}\right)^{-1}+\mathbf{Z}\right)\mathbf{s}\right]+\hat{\mathbf{Q}}^{\mathcal{H}}\mathbf{n}\\
&=\sqrt{p}\left(\hat{\mathbf{R}}\mathbf{\Sigma}^{2}_{h}\left(\mathbf{\Sigma}^{2}_{\hat{h}}\right)^{-1}+\hat{\mathbf{Q}}^{\mathcal{H}}\mathbf{Z}\right)\mathbf{s}+\hat{\mathbf{Q}}^{\mathcal{H}}\mathbf{n},
\label{detectedSignal}
\end{align}
where $\hat{\mathbf{Q}}$ and $\hat{\mathbf{R}}$ are the $N\times N$ unitary matrix (with its columns representing the orthonormal ZF nulling vectors) and $N\times M$ upper triangular matrix, respectively, given $\hat{\mathbf{H}}$ (i.e., $\hat{\mathbf{H}}\triangleq \hat{\mathbf{Q}}\hat{\mathbf{R}}$).

Consequently, the received SNR of the $i^{\rm th}$ stream ($1\leq i\leq M$), defined as $\gamma_{i}$, is presented as
\begin{align}
\gamma_{i}=p\left|\left(\hat{\mathbf{R}}\mathbf{\Sigma}^{2}_{h}\left(\mathbf{\Sigma}^{2}_{\hat{h}}\right)^{-1}+\hat{\mathbf{Q}}^{\mathcal{H}}\mathbf{Z}\right)\right|^{2}_{i,i}.
\label{SNR}
\end{align}
Unlike most related previous works thus far (e.g., see \cite{j:WangMurch2007,j:MatthaiouZFMassiveMIMO2013} and references therein), the channel estimation error term within \eqref{detectedSignal} is treated as a \emph{signal rather as noise or interference} since it can be typically demodulated via envelope detection \cite{j:LiLin2016}; hence, it may further boost the received SNR. To this end, the resultant SNR expression in \eqref{SNR} is rigorous and manifests a clear physical meaning.  

\section{Performance Metrics}

\subsection{Outage Probability}
Given $\hat{\mathbf{H}}$ (and thus $\hat{\mathbf{R}}$), the SNR expression in \eqref{SNR} introduces a non-central chi-squared RV, namely, $\gamma_{i}\overset{\text{d}}=\mathcal{X}^{2}_{2}(\mu_{i}|\hat{r}_{i,i}|^{2},\sigma^{2}_{i})$, where $\mu_{i}=p (\sigma^{2}_{h_{i}}/\sigma^{2}_{\hat{h}_{i}})^{2}$ and $\sigma^{2}_{i}=p \sigma^{2}_{Z_{i}}/2$, represent the non-centrality parameter and variance of each DoF, respectively.
This is due to the fact that $\gamma_{i}$ can be rewritten as $\gamma_{i}\triangleq |T_{i}|^{2}$, where $T_{i}=\sqrt{p}(\hat{\mathbf{R}}\mathbf{\Sigma}^{2}_{h}(\mathbf{\Sigma}^{2}_{\hat{h}})^{-1}+\hat{\mathbf{Q}}^{\mathcal{H}}\mathbf{Z})_{i,i}$. Doing so, $T_{i}\overset{\text{d}}=\mathcal{CN}(\mu_{i}|\hat{r}_{i,i}|^{2},2 \sigma^{2}_{i})$, while ${\rm Re}\{T_{i}\}\overset{\text{d}}={\rm Im}\{T_{i}\}\overset{\text{d}}=\mathcal{N}(\mu_{i}|\hat{r}_{i,i}|^{2},\sigma^{2}_{i})$. Thus, $\gamma_{i}={\rm Re}\{T_{i}\}^{2}+{\rm Im}\{T_{i}\}^{2}\overset{\text{d}}=\mathcal{X}^{2}_{2}(\mu_{i}|\hat{r}_{i,i}|^{2},\sigma^{2}_{i})$.
Also, $\hat{r}_{i,j}$ is the coefficient at the $i^{\rm th}$ row and $j^{\rm th}$ column of $\hat{\mathbf{R}}$. 

The conditional CDF/PDF of $\gamma_{i}$ are, respectively, given by
\begin{align}
F_{\gamma_{i}||\hat{r}_{i,i}|^{2}}(x)=1-Q_{1}\left(\sqrt{\frac{\mu_{i}|\hat{r}_{i,i}|^{2}}{\sigma^{2}_{i}}},\sqrt{\frac{x}{\sigma^{2}_{i}}}\right),
\label{cdfg}
\end{align}
and
\begin{align}
f_{\gamma_{i}||\hat{r}_{i,i}|^{2}}(x)=\frac{\exp\left(-\frac{(\mu_{i}|\hat{r}_{i,i}|^{2}+x)}{2 \sigma^{2}_{i}}\right)}{2 \sigma^{2}_{i}}I_{0}\left(\frac{\sqrt{\mu_{i}|\hat{r}_{i,i}|^{2}x}}{\sigma^{2}_{i}}\right).
\label{pdfg}
\end{align}
The unconditional CDF of $\gamma_{i}$ is given by $F_{\gamma_{i}}(x)=\int^{+\infty}_{0}F_{\gamma_{i}||\hat{r}_{i,i}|^{2}}(x|y)f_{|\hat{r}_{i,i}|^{2}}(y)dy$ with \cite{j:GOREHeathPaulraj}
\begin{align}
f_{|\hat{r}_{i,i}|^{2}}(y)=\frac{y^{N-M}\exp\left(-y/\sigma^{2}_{\hat{h}_{i}}\right)}{(N-M)!\left(\sigma^{2}_{\hat{h}_{i}}\right)^{N-M+1}}.
\label{pdfr}
\end{align}
Thereby, utilizing \cite[Eq. (12)]{j:sofotasiosmarcum} and after some straightforward manipulations, the CDF of $\gamma_{i}$ is expressed in a closed form as
\begin{align}
\nonumber
F_{\gamma_{i}}(x)=&1-\exp\left(-\frac{(\sigma^{2}_{h_{i}}+1/p)x}{\sigma^{2}_{h_{i}}}\right)\\
&\times\Bigg[1+x\sum^{N-M}_{l=0}\left(\frac{1}{p \sigma^{2}_{h_{i}}+1}\right)^{l}{}_1F_{1}\left(l+1;2;x\right)\Bigg].
\label{cdfUncond}
\end{align}
It is noteworthy to state here that the hypergeometric series in \eqref{cdfUncond} can be further relaxed to a finite sum series according to \cite[Eqs. (07.20.03.0026.01) and (07.20.03.0025.01)]{wolfram}, including elementary-only functions due to their involved integer-valued parameters, yielding
\begin{align}
\nonumber
F_{\gamma_{i}}&(x)=1-\exp\left(-\frac{x}{p \sigma^{2}_{h_{i}}}\right)\\
&\times\Bigg[1+\sum^{N-M}_{l=1}\sum^{l-1}_{k=0}\left(\frac{1}{p \sigma^{2}_{h_{i}}+1}\right)^{l}\frac{(1-l)_{k} (-1)^{k}x^{k+1}}{k!(2)_{k}}\Bigg].
\label{cdfUncondd}
\end{align}
Hence, $F_{\gamma_{i}}(\cdot)$ can be easily, accurately and rapidly computed, especially in the case when $N\gg M\gg 1$. Further, it corresponds to the exact outage probability, say $P_{\rm out}(\cdot)$, since $P_{\rm out}(\gamma^{(i)}_{\rm th})=F_{\gamma_{i}}(\gamma^{(i)}_{\rm th})$ for a predetermined SNR threshold $\gamma^{(i)}_{\rm th}\triangleq \exp(\mathcal{R}_{i})-1$, where $\mathcal{R}_{i}$ denotes the target rate of the $i^{\rm th}$ transmitted stream in nats per channel use (npcu).

Additionally, noticing from \eqref{SNR} that $\hat{\mathbf{H}}$ and $\mathbf{Z}$ are mutually independent and $\mathbf{Z}$ is zero-mean, the average received SNR of the $i^{\rm th}$ stream, $\overline{\gamma}_{i}$, is given by 
\begin{align}
\nonumber
\overline{\gamma}_{i}&\triangleq \mathbb{E}\left[p\left|\left(\hat{\mathbf{R}}\mathbf{\Sigma}^{2}_{h}\left(\mathbf{\Sigma}^{2}_{\hat{h}}\right)^{-1}+\hat{\mathbf{Q}}^{\mathcal{H}}\mathbf{Z}\right)\right|^{2}_{i,i}\right]\\
&=\frac{p (N-M+1) \sigma^{4}_{h_{i}}}{\sigma^{2}_{h_{i}}+\frac{1}{p}}+\frac{\sigma^{2}_{h_{i}}}{p \sigma^{2}_{h_{i}}+1}.
\label{avgSNR}
\end{align}
Obviously, in the very high SNR regime (i.e., when $p\rightarrow +\infty$), the impact of channel estimation error vanishes, as expected, and $\overline{\gamma}^{(p\rightarrow +\infty)}_{i}\rightarrow p \sigma^{2}_{h_{i}}(N-M+1)$.

\subsection{Design of the Efficient Rate $\mathcal{R}_{i}$}
Unlike the conventional ergodic setting (i.e., assuming an enormously high number of channel uses per each block/frame), short packet communications require a careful design of the appropriate $\mathcal{R}_{i}$ since they operate, in principle, on non-ergodic channels. Capitalizing on the properties of the considered QR decomposition, the total detected signal in \eqref{detectedSignal} can be viewed as $M$ parallel single input-multiple output (SIMO) signals in AWGN channels. Thereby, there is a tight approximation that interrelates the channel coding rate $\mathcal{R}_{i}$, SNR, number of channel uses for information transmission per frame $L$, and the target (i.e., maximum allowable) frame error rate $\epsilon$, which is defined as \cite[Eq. (4)]{j:YangDurisi2014}
\begin{align}
\mathcal{R}^{\star}_{i}&=\underset{\mathcal{R}_{i}}{{\rm arg}}\left\{F_{\gamma_{i}}\left(\exp(\mathcal{R}_{i})-1\right)=\epsilon \right\}+\mathcal{O}\left({\rm ln}(L)/L\right),
\label{optRate}
\end{align}
where $\mathcal{R}^{\star}_{i}$ denotes the maximum rate of the $i^{\rm th}$ stream satisfying a target error probability $\epsilon$ and using a total blocklength of $L$ channel uses, given a fixed set of $\{N,M,p,\sum^{M}_{i=1}\sigma^{2}_{\hat{h}_{i}}\}$ values. The arising mismatch between \eqref{optRate} and the true rate, caused by truncating the higher order terms $\mathcal{O}({\rm ln}(L)/L)$, can be tightly approximated by the following error probability\cite[Eq. (95)]{j:YangDurisi2014}  
\begin{align}
P_{\rm err}(\mathcal{R}^{\star}_{i})=\mathbb{E}\left[Q\left(\frac{\sqrt{L}({\rm ln}(1+\gamma_{i})-\mathcal{R}^{\star}_{i})}{\sqrt{1-(1+\gamma_{i})^{-2}}}\right)\right],
\label{finiteblocklengthApprox}
\end{align}
where the latter expectation is evaluated with respect to $\gamma_{i}$. Also, \eqref{finiteblocklengthApprox} is efficient for $L\geq 100$, while it tends to coincide with the outage probability, i.e., $P_{\rm err}(\mathcal{R}^{\star}_{i})\rightarrow F_{\gamma_{i}}(\exp(\mathcal{R}^{\star}_{i})-1)$ for increasing $L$. Finally, by inserting \eqref{cdfUncondd} into \eqref{optRate}, $\mathcal{R}^{\star}_{i}$, can be numerically computed quite easily for an arbitrary number of the transceiver antennas.

\subsection{Effective Number of Transmitted Streams $M$}
For a given channel coding rate $\mathcal{R}_{i}\:\:\forall i$ and a total blocklength $L$, we aim to maximize the effective total goodput (in npcu) so as to derive the appropriate number of $M$, say $M^{\star}$. The total system goodput is defined as \cite{j:MakkiSvensson2014,j:Schiessl2016}
\begin{align}
\nonumber
G&\triangleq \left(1-\frac{M}{L_{\rm T}}\right)\sum^{M}_{i=1}\mathcal{R}_{i}[1-P_{\rm err}(\mathcal{R}_{i})]\\
&\approx \left(1-\frac{M}{L_{\rm T}}\right)\sum^{M}_{i=1}\mathcal{R}_{i}[1-P_{\rm out}(e^{\mathcal{R}_{i}}-1)],
\label{goodputgen}
\end{align}
which takes into account the overhead (with regards to the actual data rate) caused by the channel uses utilized for the CSI training phase. Thus, $M^{\star}$ arises as the solution of the following optimization problem:
\begin{align}
\nonumber
&\underset{M}{{\rm arg}\max} \quad G\\
&{\rm s.t.} \quad 0\leq M\leq N.
\label{optProblem}
\end{align}
Unfortunately, the above problem is non-convex in general since the optimization variable is placed at the upper sum limit of the objective function. However, it can be quite easily computed numerically as a simple line search over the integers with a related computational complexity which is (at most) in the order of $\mathcal{O}(N)$. For completeness of exposition, the proposed iterative approach is formalized in Algorithm~1.

\begin{algorithm}[t]
	\caption{Effective \# of Transmitted Streams $M$}
	\begin{algorithmic}[1]
		 \INPUT{$N$, $M$, $p$, $\{\mathcal{R}_{i},\sigma^{2}_{h_{i}}\}^{M}_{i=1}$}
		 \OUTPUT{$M^{\star}$ (the effective number of transmitted streams)}
		  \WHILE{$M>0$}
			\STATE{Sort the $M$ streams from the highest to lowest priority. In the case of identical priority per stream, sort the $M$ streams such that $\mathcal{R}_{M}[1-P_{\rm out}(e^{\mathcal{R}_{M}}-1)]\geq \cdots \geq \mathcal{R}_{1}[1-P_{\rm out}(e^{\mathcal{R}_{1}}-1)]$}
			\FOR{$j=M:-1:2$}
			\STATE{Compute Eq. \eqref{goodputgen} given $j$, namely $G(j)$;}
			\STATE{Compute Eq. \eqref{goodputgen} given $j-1$, namely $G(j-1)$;}
				\IF{$G(j)\geq G(j-1)$}
					\STATE $M^{\star} = j$;
					\STATE End of the algorithm;
					\ELSE $\:\:j=j-1$
					\STATE Go to Step 4;
				\ENDIF
			\ENDFOR
		\ENDWHILE 
	\end{algorithmic}
\end{algorithm} 

Motivated by the above statement, we retain our focus on the case of large yet finite $N$ and/or $M$ (i.e., the so-called massive MIMO regime), and identical statistics where $\sigma^{2}_{h_{i}}\triangleq \sigma^{2}_{h}\:\:\forall i$ and $\mathcal{R}_{i}\triangleq \mathcal{R}\:\:\forall i$. This scenario of identical statistics can find direct application in transmitters with co-located antennas and/or when the large-scale fading amongst the involved transmitters and the intended receiver at each resource block is quite similar (e.g., via location-driven user scheduling). In this case, the total goodput becomes 
\begin{align}
G\triangleq \left(1-\frac{M}{L_{\rm T}}\right)M \mathcal{R}[1-P_{\rm out}(e^{\mathcal{R}}-1)],
\end{align}
while
\begin{align}
\nonumber
G&=\left(1-\frac{M}{L_{\rm T}}\right)M \mathcal{R}\exp\left(-\frac{(\sigma^{2}_{h_{i}}+1/p)(e^{\mathcal{R}}-1)}{\sigma^{2}_{h_{i}}}\right)\\
\nonumber
&\ \ \ \ \times\Bigg[1+\sum^{N-M}_{l=0}\frac{(e^{\mathcal{R}}-1)}{\left(p \sigma^{2}_{h_{i}}+1\right)^{l}}\:{}_1F_{1}\left(l+1;2;e^{\mathcal{R}}-1\right)\Bigg]\\
\nonumber
&\geq G_{\rm LB}\triangleq \left(1-\frac{M}{L_{\rm T}}\right)M \mathcal{R}\exp\left(-\frac{(\sigma^{2}_{h}+1/p)(e^{\mathcal{R}}-1)}{\sigma^{2}_{h}}\right)\\
&\ \ \ \ \times\Bigg[1+\sum^{N-M}_{l=0}\frac{(e^{\mathcal{R}}-1)}{\left(p \sigma^{2}_{h_{i}}+1\right)^{l}}\left(1+\frac{(l+1) (e^{\mathcal{R}}-1)}{2}\right)\Bigg],
\label{goodputBound}
\end{align} 
where the latter lower bound on the total goodput, defined as $G_{\rm LB}$, is obtained by utilizing \eqref{cdfUncond} and using the inequality \cite[Eq. (1.2)]{j:Shehata2015} ${}_1F_{1}(a;2;x)\geq 1+a x/2$, which is quite tight (as illustrated in the next section) when $\mathcal{R}$ is relatively low (e.g., $\mathcal{R}\leq 0.1$), as in the ultra-reliable region.\footnote{Ultra-reliable communications require extremely low packet error probability (usually $\ll 10^{-5}$); thereby, reflect on very robust channel coding schemes and correspondingly on rather low coding rates.} Most importantly, the sum-series term within \eqref{goodputBound} can be conveniently expanded after some straightforward calculus as
\begin{align}
\nonumber
&\sum_{l=0}^{N-M} A^l (1 + (l + 1) B)=(A-1)^{-2} \big\{1+A+B \big[A^{N-M}\\
&\times [A (1 + B + B (N-M))-B(2+N-M)-1]-1\big]\big\},
\label{SumExpansion}
\end{align}
where $A\triangleq 1/(p \sigma^{2}_{h}+1)$ and $B\triangleq (e^{\mathcal{R}}-1)/2$ are introduced for presentation clarity.

Unlike the actual goodput, its lower bound in \eqref{goodputBound}, $G_{\rm LB}$, is indeed a concave function. This can be easily verified by inserting \eqref{SumExpansion} into \eqref{goodputBound} and showing that the resultant second derivative with respect to $M$ is a negative function. Thereby, $M^{\star}$ is lower bounded by $M^{\star}\geq \min\{\left\lfloor J\right\rfloor,N\}$, where $J$ denotes the numerical solution of $\frac{{\rm d}}{{\rm d}M}G_{\rm LB}=0$ with respect to $M$. This approach is rather simple and more computationally efficient than the iterative method previously proposed to solve \eqref{optProblem}; especially when the receiver is equipped with a vast antenna array. 

\section{Numerical Results and Discussion}
In this section, the derived analytical results are verified via numerical validation, whereas they are cross-compared with corresponding Monte-Carlo simulations. Also, without loss of generality and for the sake of clarity, hereinafter, we assume an identical statistical profile for each transmitted stream (i.e., $\sigma^{2}_{h_{i}}\triangleq \sigma^{2}_{h}=0.1$ and $\mathcal{R}_{i}\triangleq \mathcal{R}\:\:\forall i$) in order to evaluate more concretely the overall system performance (on average) and obtain more impactful insights. Subsequently, the convention used to denote the Tx-Rx antenna array is in the form of $M\times N$. Also, line-curves and solid dot-marks denote the analytical and simulation results, respectively.

\begin{figure}[!t]
\centering
\includegraphics[trim=2.0cm 0.2cm 0.5cm .6cm, clip=true,totalheight=0.25\textheight]{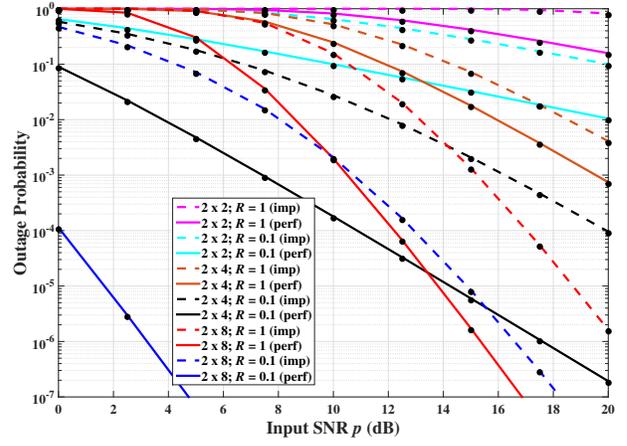}
\caption{Outage probability vs. various input SNR values and system setups when $L=300$.}
\label{fig1}
\end{figure}

In Fig.~\ref{fig1}, the outage probability per stream is illustrated for various system setups, when $M=2$ and for a total blocklength of $L_{\rm T}=L+M=302$ channel uses. The exact simulation results are given by numerically evaluating \eqref{finiteblocklengthApprox}, which sharply match the derived analytical results, which are computed via \eqref{cdfUncondd}. Abbreviations `imp' and `perf' in Fig.~\ref{fig1} stand for the practical imperfect CSI and ideal perfect CSI case, respectively. The latter case is computed by the well-known formula $P_{\rm out}(x)=1-\Gamma(N-M+1,x/(p \sigma^{2}_{h}))/\Gamma(N-M+1)$. Obviously, the system performance is enhanced for higher $N$ and/or lower coding rate $\mathcal{R}$, as expected. Moreover, the performance gap between the ideal CSI and imperfect CSI scenarios is quite high and should be regarded in practical system settings. Interestingly, the said performance gap increases for a higher receive antenna array $N$ and/or a lower coding rate $\mathcal{R}$ (which is usually the case in the ultra-reliable region \cite{j:AnandVeciana2018}).
 
\begin{figure}[!t]
\centering
\includegraphics[trim=1.5cm 0.2cm 0.5cm .6cm, clip=true,totalheight=0.25\textheight]{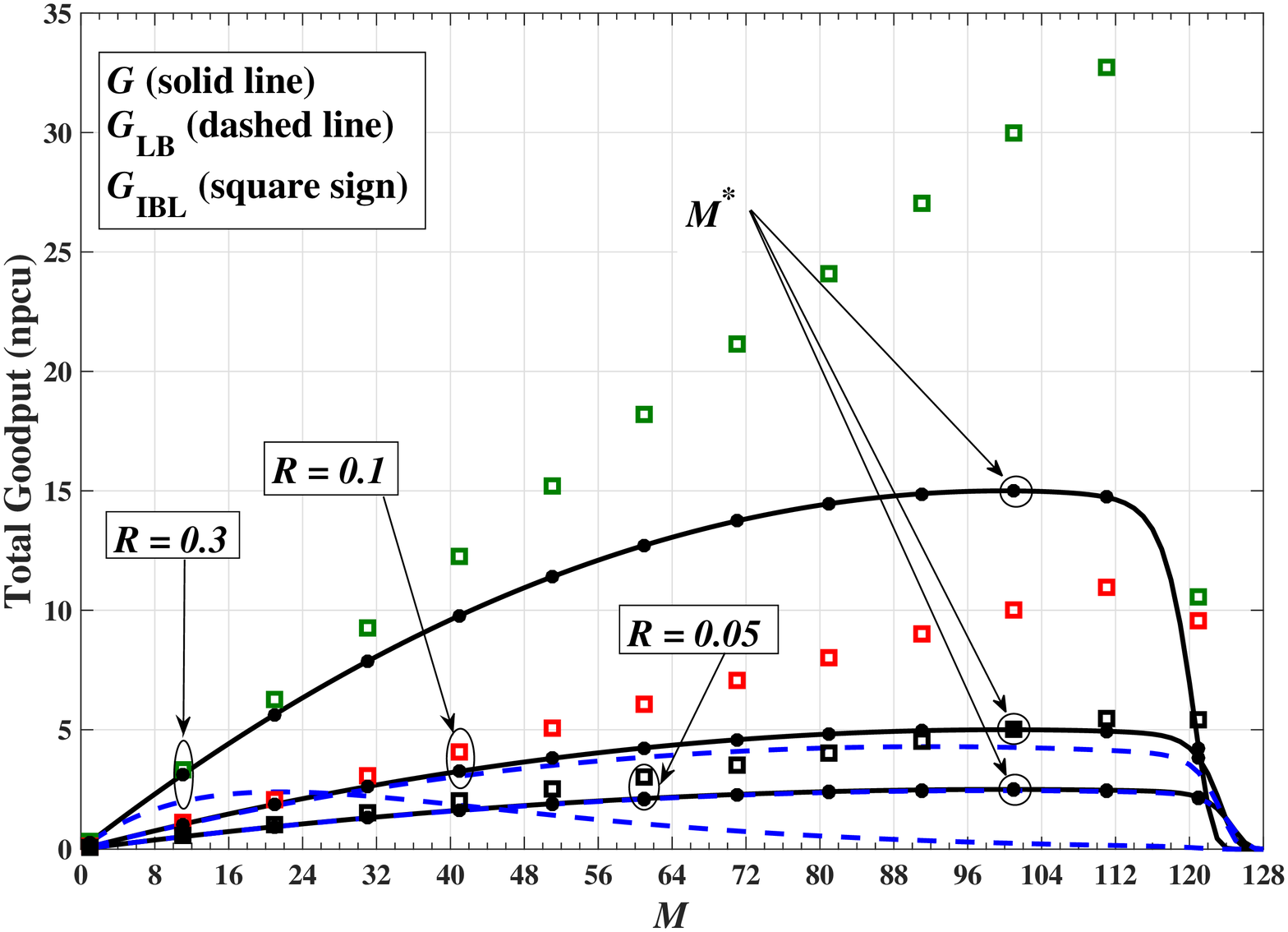}
\caption{Total system goodput vs. various $M(\leq N)$ values and channel coding rates when $L=200$, $N=128$, $P_{\max}=20$dB, and $p=P_{\max}/\sqrt{N}\approx 9.46$dB. Also, $M^{\star}$ is obtained as per Algorithm~1, while $G_{\rm IBL}$ stands for the total goodput when $L=10^{4}$ (i.e., the conventional infinite blocklength).}
\label{fig3}
\end{figure}

Finally, Fig.~\ref{fig3} demonstrates the total system goodput when a large antenna array is placed at the receiver. Further, the energy efficient transmission (i.e., green communication) is considered herein, where the transmit power is proportionally reduced with regards to $N$. In this particular illustrative example, and without loss of generality, a power scaling low of $p\triangleq P_{\max}/\sqrt{N}$ is adopted with $P_{\max}$ denoting the maximum achievable transmit power \cite{j:Zhang2014}. This strategy can be easily established, e.g., via an open-loop power control. As mentioned earlier, the lower bound on the total goodput, as per \eqref{goodputBound}, tightly approximates the actual goodput for values of coding rate $\mathcal{R}\leq 0.1$, while it is quite loose for higher values. Most importantly, the accuracy of the proposed iterative approach is verified by providing the effective number of transmitted streams $M^{\star}$ (as per Algorithm~1) in the short-packet transmission regime. It is obvious that $M^{\star}$ indicates the case when the total goodput is being maximized within the range $1\leq M \leq N$. Notably, the maximization of total goodput for short-packet (i.e., finite blocklength) communications requires a different number of simultaneously transmitted streams, $M$, in comparison to the conventional infinite blocklength communications.

\section{Concluding Remarks}
A multiuser MIMO wireless communication system was analytically studied under the short-packet transmission regime; a suitable approach for the forthcoming URLLC services. Under independent Rayleigh faded channels and imperfect CSI at the receiver, the spatial multiplexing mode of transmit operation is adopted along with ZF detection. Unlike most previous works, we proposed an approach where the channel estimation error is treated as a signal rather than noise. Doing so, new and exact closed-form expressions were derived for some key performance metrics, i.e., the outage probability and system goodput. Also, a lower bound of the system goodput was derived for the simplified case of identical channel fading conditions. Moreover, an iterative algorithm was formulated to define the effective number of simultaneously transmitted streams so as to enhance the system goodput. Finally, some useful engineering insights have been manifested, such as the impact of channel estimation onto the system performance; the gain obtained from multiple antennas; the key role of coding rate; and the emphatic performance difference between the finite (i.e., short-packet) and infinite blocklength transmission.

\bibliographystyle{IEEEtran}
\bibliography{IEEEabrv,References}

\vfill

\end{document}